\documentclass[aps,prl,twocolumn,groupedaddress]{revtex4}
\usepackage{amsfonts}
\usepackage{amssymb}
\usepackage{amsmath}
\usepackage{amsmath,amssymb,bm,graphicx}
\usepackage{mathrsfs}

\begin{document}

\title{ Minimizers with discontinuous velocities for the electromagnetic
variational method}
\author{Jayme De Luca}
\email[author's email address:]{ deluca@df.ufscar.br}
\affiliation{Universidade Federal de S\~{a}o Carlos, \\
Departamento de F\'{\i}sica\\
Rodovia Washington Luis, km 235\\
Caixa Postal 676, S\~{a}o Carlos, S\~{a}o Paulo 13565-905\\
Brazil\\
}
\date{\today }

\begin{abstract}
The electromagnetic two-body problem has \emph{neutral differential delay}
equations of motion that, for generic boundary data, can have solutions with 
\emph{discontinuous} derivatives. If one wants to use these neutral
differential delay equations with \emph{arbitrary} boundary data, solutions
with discontinuous derivatives must be expected and allowed. Surprisingly,
Wheeler-Feynman electrodynamics has a boundary value variational method for
which minimizer trajectories with discontinuous derivatives are also
expected, as we show here. The variational method defines continuous
trajectories with piecewise defined velocities and accelerations, and
electromagnetic fields defined \emph{by} the Euler-Lagrange equations \emph{%
on} trajectory points. Here we use the piecewise defined minimizers with the
Li{\'{e}}nard-Wierchert formulas to define generalized electromagnetic
fields almost everywhere (but on sets of points of zero measure where the
advanced/retarded velocities and/or accelerations are discontinuous). Along
with this generalization we formulate the \emph{generalized absorber
hypothesis} that the far fields vanish asymptotically \emph{almost everywhere%
} and show that localized orbits with far fields vanishing almost everywhere 
\emph{must} have discontinuous velocities on sewing chains of breaking
points. We give the general solution for localized orbits with vanishing far
fields by solving a (linear) neutral differential delay equation for these
far fields. We discuss the physics of orbits with discontinuous derivatives
stressing the differences to the variational methods of classical mechanics
and the existence of a spinorial four-current associated with the
generalized variational electrodynamics.
\end{abstract}

\pacs{05.45.-a, 02.30.Ks, 03.50.De,41.60.-m}
\maketitle

\section{Introduction}

Non-radiating motion of extended charge distributions in classical
electrodynamics has been known to exist for some time (c.f. \cite{1,2,3,4}
and references therein, and \cite{5,6,7}). On the other hand, for systems
with a few \emph{point charges}, Larmor's radiation of energy at a rate
proportional to the squared modulus of the acceleration plagues classical
electrodynamics. To construct orbits that do not radiate, and hence are
without acceleration, a simple option are constant velocity motions, which
imply unbounded motion.

Along \emph{bounded} two body motions supported by mutual action at a
distance, we expect acceleration to be needed to change velocities, unless
velocities are allowed to change discontinuously. For example, periodic
polygonal orbits with piecewise constant velocity segments have vanishing
radiation fields.

Here we extend Wheeler-Feynman electrodynamics \cite{Fey-Whe} to include
motion with discontinuous velocities. This is a natural extension provided
by the variational boundary value problem \cite{JMP2009}. The resulting
extended electrodynamics has several appealing physical features: (i) There
exists a scalar function (the finite action \cite{JMP2009}), and the
condition for a minimizer demands that the partial derivatives of the
action, with respect to each particle's four-velocity, be continuous along
minimal orbits. These continuous four-component linear currents are
analogous to the Dirac-equation of quantum mechanics, thus endowing the
extended Wheeler-Feynman electrodynamics with spin. This is a feature not
present in any other classical electrodynamics of point-charges; (ii)
Besides naturally including non-radiating orbits, the extended
electrodynamics can be shown to lead simply to a de Broglie length for
double-slit scattering upon detailed modeling \cite{double-slit}; (iii) The
absorber hypothesis, first idealized to hold as an average over an infinite
universe\cite{Fey-Whe}, has no known solutions \cite{Martinez} for many-body
motion in Wheeler-Feynman theory \cite{Martinez, Marino, Schoenberg, Schild,
vonBaeyer} with which it is consistent. Here we show that the variational
electrodynamics allows a concrete realization of the absorber hypothesis for
a two-particle universe, i.e., there exists a non-empty class of two-body
motions with vanishing far-fields, so that we do not need either large
universes or randomization \cite{Boeyer, Nelson}; and (iv) two-body orbits
with vanishing far-fields were used in Ref. \cite{stiff-hydrogen} to predict
spectroscopic lines for hydrogen with a few percent precision. 

Since the speed of light is constant in inertial frames, the equations of
motion for point-charges are state dependent differential delay equations.
More specifically, Wheeler-Feynman electrodynamics \cite{Fey-Whe,
double-slit, Dirk} has mixed-type state-dependent neutral differential delay
equations of motion for the two-body problem.

The theory of delay equations is still incomplete \cite%
{BellenZennaro,Hans-Otto} but it is known that purely-retarded differential
delay equations with generic $C^{1}$ initial histories have continuous
solutions with a discontinuous derivative at the initial time. The
derivative becomes continuous at the next breaking point \cite{BellenZennaro}
and progresses from $C^{k}$ to $C^{k+1}$at successive breaking points. On
the other hand, a purely retarded neutral differential delay equation with a
generic $C^{1}$ initial history \cite{BellenZennaro} can have continuous
solutions with discontinuous derivatives at \emph{all} breaking points.

If one wants to use the electromagnetic neutral differential delay equations
with arbitrary boundary data, solutions with discontinuous derivatives must
be expected and accommodated. Surprisingly, this same neutrality is
compatible with the recently developed boundary-value-variational method for
Wheeler-Feynman electrodynamics \cite{JMP2009}. For orbits where the
acceleration is not defined at a few points, the variational method offers a
well-posed alternative to define trajectories beyond those satisfying a
Newtonian-like neutral differential delay equation \emph{everywhere}. The
variational method involves an integral that requires only piecewise-defined
velocities, generalizing naturally to continuous orbits with discontinuous
derivatives at breaking points.

Our generalized electrodynamics contains the $C^{2}$ orbits of the
Wheeler-Feynman theory. As shown in Ref. \cite{JMP2009}, if boundary data
are such that the extremum orbit is piecewise $C^{2}$ with \emph{continuous}
velocities, the Wheeler-Feynman equations hold everywhere with the exception
of a countable set of points where accelerations are discontinuous (which is
a set of measure zero for the action integral). We henceforth define a
breaking point as a point where velocity or acceleration are discontinuous.
Here we show that continuous orbits with discontinuous velocities are
possible minimizers if these satisfy further continuity-conditions. These
continuity conditions are non-local, unlike the conditions for an extremum
of the variational methods of classical mechanics, which do \emph{not} allow
discontinuous velocities. Finally, if the extremum is not piecewise $C^{2}$,
the variational method defines minimizers that are not described by
piecewise-defined-Wheeler-Feynman neutral differential delay equations
(which are not studied here).

To discuss the relationship to Maxwell's electrodynamics it is important to
keep in mind that: (i) Wheeler-Feynman electrodynamics is a theory of \emph{%
trajectories}, where fields are only \emph{derived quantities}; and (ii) the
boundary-value-variational-method defines only a \emph{finite} segment of a
trajectory, rather than a global trajectory \cite{JMP2009}. The variational
equations along piecewise $C^{2}$ orbits include the electromagnetic fields
in the Euler-Lagrange equations \cite{JMP2009}, which are used here to give
a derived operational meaning to the electromagnetic fields \cite{Darryl}.
The electromagnetic fields appear as coupling terms of the variational
equations and are defined \emph{on} trajectory segments by the usual
electromagnetic formulae \cite{JMP2009}.

In our generalization we use the Li\'{e}nard-Wierchert electromagnetic
formulae to define fields by extension at all space-time-points for which
future and past lightcones fall in the finite segment of the minimizer
trajectory. For continuous trajectories with discontinuous velocities,
and/or accelerations on sets of measure zero, we construct the
electromagnetic fields only for points having a future and past lightcone,
leaving the fields undefined where the past or future lightcones have a
discontinuous velocity/acceleration (usually another set of measure zero).
We further introduce the concept of short-range orbits as localized orbits
with far-fields vanishing almost everywhere. This bears a close relation to
the electromagnetic notion of radiation \cite{Darryl}. \ 

In their original articles, Wheeler and Feynman \cite{Fey-Whe} attempted to
derive an electrodynamics with retarded-only fields from the hypothesis that
the universal far-fields vanish at all times (the absorber hypothesis) \cite%
{Fey-Whe}. Here we generalize the absorber hypothesis \cite{Fey-Whe} to
include fields that can be undefined on sets of measure zero, thus arriving
at the \emph{generalized absorber hypothesis} (GAH) that the far-fields
vanish \emph{almost everywhere}. We show that short-range-two-body-orbits 
\emph{must} involve discontinuous derivatives on a countable set of points.

One advantage of our generalization is to include
spatially-bounded-globally-defined-continuous-orbits with far-fields
vanishing almost everywhere, which we call short-range orbits. This
generalization presents itself naturally as the next option after one shows
that there are no $C^{2}$ localized orbits with far-fields vanishing
everywhere. The short-range piecewise $C^{2}$ continuous orbits are
naturally described by the variational method in the same way as the
globally $C^{2}$ continuous orbits. However, the former are minimizers
inside a larger family of boundary-data, which is the second advantage of
our generalization. This extended class of orbits includes orbits that are
limits of Cauchy sequences of orbits with far-fields disturbing the universe
less and less, i.e., the vanishing-far-field-limit of the GAH.

In this paper we use the word "minimizer" meaning a generalized critical
point of the variational method, that could be either a minimum or a saddle
point. The paper is divided as follows: In Section II we discuss the
variational method for piecewise-defined continuous orbits with
discontinuous derivatives. We show that the variational method prescribes a
continuous momentum current at each breaking point in addition to
Euler-Lagrange equations from each side of the breaking point. We discuss
how the non-local-momentum-currents can be conserved even in the presence of
velocity discontinuities along ``sewing chains". In Section III we prove
that globally-defined short-range bounded orbits must have discontinuous
velocities on a sewing chain of breaking points by giving the general
solution to a neutral differential delay equation for the far-fields. In
Section IV we discuss the physics of generalized minimizers along with some
open questions and differences from bounded orbits to unbounded scattering
orbits.

\section{Boundary value variational method}

The variational method \cite{JMP2009} is well defined for continuous
trajectories $\boldsymbol{x}_{1}(t_{1})$ and $\boldsymbol{x}_{2}(t_{2})$ $%
\in 
\mathbb{R}
^{3}$ that are piecewise $C^{1}$. The boundary conditions for the
variational method \cite{JMP2009} are illustrated in Figure 1, i.e., the
initial point $O_{A}$ for trajectory $1$ plus the segment of trajectory $2$
inside the lightcone of $O_{A}$, and \ the endpoint $L_{B}$ for the
trajectory $2$ plus the segment of trajectory of particle $1$ inside the
lightcone of $L_{B}$. \ For variations of trajectory $1$ the action
functional \cite{JMP2009} reduces to 
\begin{eqnarray}
S &\equiv &K_{2}+\int_{0}^{T_{L^{-}}}\mathscr{L}(\mathbf{x}_{1},\mathbf{v}%
_{1},\mathbf{x}_{2},\mathbf{v}_{2})dt_{1}  \label{action1} \\
&=&-\int_{0}^{T_{L^{-}}}m_{1}\sqrt{1-\mathbf{v}_{1}^{2}}dt_{1}  \notag \\
&&+\int_{0}^{T_{L^{-}}}\frac{(1-\mathbf{v}_{1}\cdot \mathbf{v}_{2+})}{%
2r_{12+}(1+\mathbf{n}_{12+}\mathbf{\cdot v}_{2+})}dt_{1}  \notag \\
&&+\int_{0}^{T_{L^{-}}}\frac{(1-\mathbf{v}_{1}\cdot \mathbf{v}_{2-})}{%
2r_{12-}(1-\mathbf{n}_{12-}\mathbf{\cdot v}_{2-})}dt_{1},  \notag
\end{eqnarray}%
where $K_{2}$ depends only on trajectory $2$ and quantities of particle $2$
are defined at times $t_{2\pm}(\mathbf{x}_1 (t_1))$ according to the implicit condition for the
advanced/retarded light cones of $t_{1}$, i.e.,%
\begin{equation}
t_{2\pm }(\mathbf{x}_1 (t_1)) =t_{1}\pm |\mathbf{x}_{1}(t_{1})-\mathbf{x}_{2}(t_{2\pm })|.
\label{lightcone}
\end{equation}%
In Eq. (\ref{action1}) the $r_{12+}\equiv |\mathbf{x}_{1}(t_{1})-\mathbf{x}%
_{2}(t_{2\pm })|$ are the distances from $\mathbf{x}_{1}(t_{1})$ to\ the
respective advanced/retarded position $\mathbf{x}_{2}(t_{2\pm })$ along
trajectory $2$, unit vector $\mathbf{n}_{12\pm }$ points from $\mathbf{x}%
_{1}(t_{1})$ to the respective advanced/retarded position $\mathbf{x}%
_{2}(t_{2\pm })$, i.e., $\mathbf{n}_{12\pm }\equiv $ $(\mathbf{x}_{1}(t_{1})-%
\mathbf{x}_{2}(t_{2\pm }))/r_{12\pm }$ and last $\mathbf{v}_{2\pm }\equiv d%
\mathbf{x}_{2}/dt_{2}|_{t_{2\pm }}$. \ Notice that Eq. (\ref{action1}) is an
integral over the velocities, and is a well-defined operation even for
trajectories with discontinuous velocities (and even more general types of
continuous trajectories with square-integrable velocities that are not
studied here).

Here we extend Wheeler-Feynman electrodynamics to trajectories with
discontinuous velocities on a countable set of points using the
boundary-value-variational-method. For piecewise $C^{1}$ trajectories (\emph{%
and} piecewise $C^{1}$ histories) it is possible to define disjoint
intervals $t\in $ $[l_{\sigma -1}^{+},l_{\sigma }^{-}]$, with $l_{\sigma
}^{-}=l_{\sigma }^{+}$, $\ $for $\sigma =1,...,N$, where the continuous
trajectory $\boldsymbol{x}_{1}(t_{1})$ and delayed arguments $t_{2\pm
}(t_{1})$ are piecewise $C^{1}$. The upper plus in $l_{\sigma }^{+}$
indicates the right-limit of the $\sigma^{ th}$ breaking point while the
upper minus in $l_{\sigma }^{-}$ indicates the left-limit of the $\sigma
^{th}$ breaking point. These are not to be confused with the lower indices
used to denote quantities evolving in future and past light cones. 
\begin{figure}[h]
\centering
\includegraphics[scale=0.47]{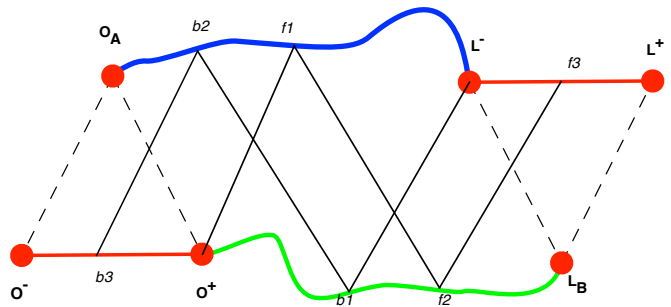} 
\caption{ Illustrated in red is the initial point $O_{A}$ of trajectory $1$
plus the segment of trajectory $2$ inside the lightcone of $O_{A}$, i.e.,
from point $O^{-}$ to point $O^{+}$ and \ the endpoint $L_{B}$ of trajectory 
$2$ plus the segment of trajectory of particle $1$ inside the lightcone of $%
L_{B}$, i.e., from point $L^{-}$ to point $L^{+}$. The trajectory of
particle $1$ of the variational method goes from $O_{A}$ to $L^{-}$ (blue
line) while the trajectory of particle $2$ goes from $O^{+}$ to $L_{B}$
(green line). The first breaking point is point $O^{+}$which generates a
forward sewing chain of breaking points $f_{1},f_{2,}f_{3}$ while endpoint $%
L^{-}$ is a breaking point generating a backward sewing chain of breaking
points $b_{1},b_{2},b_{3}$.}
\label{fig1}
\end{figure}

The variations of trajectory $1$ for the action (\ref{action1}) are defined
piecewise $C^{1}$\ with fixed endpoints, i.e.,%
\begin{eqnarray}
\mathbf{u}_{1}(t_{1}) &=&\boldsymbol{\mathbf{x}}_{1}(t_{1})+\mathbf{b}%
_{1}(t_{1}),  \label{perturb} \\
\mathbf{\dot{u}}_{1}(t_{1}) &=&\boldsymbol{\mathbf{\dot{x}}}_{1}(t_{1})+%
\mathbf{\dot{b}}_{1}(t_{1}),  \notag
\end{eqnarray}%
where and overdot denotes a time-derivative and the boundary conditions are

\begin{eqnarray}
\mathbf{b}_{1}(l_{o}^{+} &=&0)=0,  \label{boundaries} \\
\mathbf{b}_{1}(l_{N}^{-} &=&T_{L^{-}})=0.  \notag
\end{eqnarray}

If the continuous and piecewise $C^{1}$ perturbation $\mathbf{b}_{1}(t_{1})$
has a discontinuous derivative in another set of intervals $t\in $ $(h_{\mu
-1}^{+},h_{\mu }^{-})$, then the perturbed trajectory $\mathbf{u}_{1}(t_{1})$
is continuous and piecewise $C^{1}$ in the extended set of intervals defined
by all intersections of the sets $(h_{\mu -1}^{+},h_{\mu }^{-})$ and $%
(l_{\sigma -1}^{+},l_{\sigma }^{-})$. This simply increases the number of
piecewise intervals $(l_{\sigma -1}^{+},l_{\sigma }^{-})$ up to $\sigma
=M\geq N$ and the boundary condition for $\mathbf{b}_{1}(T_{L^{-}})$ of Eq. (%
\ref{boundaries}) reads 
\begin{equation}
\mathbf{b}_{1}(l_{M}^{-}=T_{L^{-}})=0.  \label{sem}
\end{equation}

Substituting the perturbed trajectory (\ref{perturb}) into the action (\ref
{action1}) and making a linear expansion about the orbit defines the Frech%
\'{e}t derivative, i.e.,

\begin{equation}
\delta S=\int_{0}^{T_{L^{-}}}[(\frac{\partial \mathscr{L} }{\partial \mathbf{%
x}_{1}}\cdot \mathbf{b}_{1})+(\frac{\partial \mathscr{L} }{\partial \mathbf{v%
}_{1}}\cdot \mathbf{\dot{b}}_{1})]dt_{1}+o(|\boldsymbol{\ b}_{1}|),
\label{Frechet}
\end{equation}%
where a ``$\cdot$" indicates the scalar product in $%
\mathbb{R}
^{3}$ and $|\boldsymbol{\ b}_{1}|$ is the sup-norm for the Banach space of
piecewise $C^{1}$ variations \cite{JMP2009}. In particular if the orbit $%
\boldsymbol{x}_{1}(t_{1}):$ $[0,T_{L^{-}}]\rightarrow 
\mathbb{R}
^{3}$ is piecewise $C^{2}$ then $\mathbf{u}_{1}(t_{1})$ is continuous and
piecewise $C^{1}$ on the same extended set of intervals.

We can integrate Eq. (\ref{Frechet}) by parts in each interval yielding%
\begin{eqnarray}
\delta S &=&\int_{0}^{T_{L^{-}}}(\boldsymbol{\mathbf{b}}_{1}\cdot \lbrack 
\frac{\partial \mathscr{L} }{\partial \mathbf{x}_{1}}-\frac{d}{dt}(\frac{%
\partial \mathscr{L} }{\partial \mathbf{v}_{1}})])dt_{1}  \label{extra} \\
&&+\sum\limits_{\sigma =1}^{\sigma =M}\int_{l_{\sigma -1}^{+}}^{l_{\sigma
}^{-}}\frac{d}{dt_{1}}(\mathbf{b}_{1}(t_{1})\cdot \frac{\partial \mathscr{L} 
}{\partial \mathbf{v}_{1}})dt_{1}.  \notag
\end{eqnarray}%
Since $\mathbf{b}_{1}(t_{1})$ is continuous we can re-arrange the second
term of the right-hand-side of Eq. (\ref{extra}) to give 
\begin{eqnarray}
\delta S &=&\int_{0}^{T_{L^{-}}}(\boldsymbol{\mathbf{b}}_{1}\cdot \lbrack 
\frac{\partial \mathscr{L} }{\partial \mathbf{x}_{1}}-\frac{d}{dt}(\frac{%
\partial \mathscr{L} }{\partial \mathbf{v}_{1}})])dt_{1}  \label{arranje} \\
&&-\sum\limits_{\sigma =1}^{\sigma =M-1}(\mathbf{b}_{1}(l_{\sigma
}^{-})\cdot \frac{\partial \mathscr{L} }{\partial \mathbf{v}_{1}}%
|_{l_{\sigma }^{-}}^{l_{\sigma }^{+}}),  \notag
\end{eqnarray}%
where 
\begin{equation}
\delta J_{1}\equiv \frac{\partial \mathscr{L} }{\partial \mathbf{v}_{1}}%
|_{l_{\sigma }^{-}}^{l_{\sigma }^{+}}=\frac{\partial \mathscr{L} }{\partial 
\mathbf{v}_{1}}(l_{\sigma }^{+})-\frac{\partial \mathscr{L} }{\partial 
\mathbf{v}_{1}}(l_{\sigma }^{-}).  \label{disc}
\end{equation}%
Equation (\ref{disc}) defines the momentum jump at $t=l_{\sigma }^{-}$,
i.e., the first (second) term on the right-hand-side of Eq. (\ref{disc}) is
the momentum evaluated from the right (left) of $t=l_{\sigma }^{-}$. 

The conditions for a critical point in the class of continuous piecewise $%
C^{1}$ orbital variations of the piecewise $C^{2}$ continuous orbit are: (i)
satisfy the Euler-Lagrange equations piecewise, to make the first term on
the right-hand-side of Eq. (\ref{arranje}) vanish; and (ii) have a
continuous momentum $\partial \mathscr{L} /\partial \mathbf{v}_{1}$ at the
breaking points so each term of the sum of the right-hand-side of Eq. (\ref%
{arranje}) vanishes for arbitrary $\mathbf{b}_{1}(l_{\sigma }^{-})$ , i.e., 
\begin{equation}
\frac{\partial \mathscr{L} }{\partial \mathbf{v}_{1}}(l_{\sigma }^{-})=\frac{%
\partial \mathscr{L} }{\partial \mathbf{v}_{1}}(l_{\sigma }^{+}).
\label{current}
\end{equation}

As is usual 
in the neighborhood of breaking points one defines derivatives from the
left-hand-side and from the right-hand-side \cite{BellenZennaro}. For the
local Lagrangians of classical mechanics one usually has $\partial %
\mathscr{L} /\partial \mathbf{v}_{1}=G_{1}(\boldsymbol{\mathbf{x}}_{1},%
\mathbf{v}_{1})$, which combined with Eq. (\ref{current}) along a continuous
trajectory would imply that \emph{each velocity} is continuous. Continuity
of velocity along a piecewise $C^{2}$ continuous orbit combined with the
Euler-Lagrange equations from each side of the breaking point further
determine a continuous acceleration, so that the orbit is actually $C^{2}$
at the breaking point. Therefore, in classical mechanics the restriction to
piecewise $C^{2}$ orbits \emph{implies} globally $C^{2}$ orbits.

However, for Eq. (\ref{action1}), or the Lagrangian given in Eq. (14) of Ref 
\cite{JMP2009}, the continuous momentum term is%
\begin{eqnarray}
\frac{\partial \mathscr{L} }{\partial \mathbf{v}_{1}} &=&\frac{m_{1}\mathbf{v%
}_{1}}{\sqrt{1-\mathbf{v}_{1}^{2}}}  \label{mv1} \\
&&-\frac{\mathbf{v}_{2-}}{2r_{12-}(1-\mathbf{n}_{12-}\cdot \mathbf{v}_{2-})}-%
\frac{\mathbf{v}_{2+}}{2r_{12+}(1+\mathbf{n}_{12+}\cdot \mathbf{v}_{2+})}, 
\notag
\end{eqnarray}%
which displays a surprising difference compared with the result obtained
from variational methods in classical mechanics.

As illustrated in Figure $1$, a simple piecewise-defined orbit has the
``sewing chain" of breaking points $(\mathit{f}_{1}\mathit{,f}_{2}\mathit{%
,\ldots }$ \,\mbox{and}\, $\mathit{b}_{1}\mathit{,b}_{2}\mathit{,\ldots }$),
where one velocity can jump \emph{if} \ the other velocity has jumped at
either the past or future breaking point. The two-body-Noether-momentum
(formula (A23) of Ref. \cite{JMP2009}), involves an integral that is
insensitive to velocity jumps plus two non-local momentum terms given by Eq.
(\ref{mv1}) (see Eqs. (A25) and (A26) of Ref. \cite{JMP2009}), that \emph{are%
} sensitive to jumps. Therefore, the two-body-Noether-momentum is conserved
as long as Eq. (\ref{mv1}) is continuous across the jumps. The first term on
the right-hand-side of Eq. (\ref{arranje}) is the Wheeler-Feynman equation
of motion for particle $1$ \cite{JMP2009} (i,.e., the usual Euler-Lagrange
equation restricted here to piecewise segments). Minimization respect to
variations of trajectory $2$ yields the neutral differential delay equation
of motion for particle $2$, and an analogous continuity with indices $1$ and 
$2$ exchanged. Notice that this surprising difference compared with the
results from variational principles in classical mechanics requires a 
\textit{minimum of two bodies and a non-local Lagrangian}.

We note that the electromagnetic variational method has a
parametrization-invariance-symmetry that allows the action (\ref{action1}) to
be expressed in Minkowski-four-space using four-velocities respect to an
arbitrary evolution parameter \cite{JMP2009}. The derivative of the
parametrization-invariant-Lagrangian with respect to the first-component of
the four-velocity 
is,%
\begin{eqnarray}
\frac{\partial \mathscr{L} }{\partial v_{1}^{o}} &=&\frac{m_{1}}{\sqrt{1-%
\mathbf{v}_{1}^{2}}}  \label{mt} \\
&&-\frac{1}{2r_{12-}(1-\mathbf{n}_{12-}\cdot \mathbf{v}_{2-})}-\frac{1}{%
2r_{12+}(1+\mathbf{n}_{12+}\cdot \mathbf{v}_{2+})}.  \notag
\end{eqnarray}%
Equation (\ref{mt}) represents the time-component of the four-momentum which
must be continuous at the breaking points of a minimizing trajectory. This
is a generalization of the argument leading to Eq. (\ref{mv1}). There is a
four-current associated with the minimization respect to each particle's
trajectory.

Last, after solving for velocity discontinuities along sewing chains of
breaking points, the second condition for a minimizer are the
piecewise-restricted Wheeler-Feynman equations of motion. These hold at each
side of a breaking point and involve the limiting accelerations from the two
sides of that breaking point. The discontinuous velocities satisfying Eqs. (%
\ref{mv1}) and (\ref{mt}), when substituted into the Wheeler-Feynman
equations at each side of a breaking point, then define a condition to be
satisfied by the acceleration discontinuities. Since this condition involves
a singular matrix \cite{JMP}, acceleration discontinuities are not fully
determined by velocity discontinuities. In general, velocity discontinuities
cause acceleration discontinuities, even though there can be special orbits
with continuous velocities and discontinuous accelerations along the null
direction of the singular matrix \cite{JMP}. Only in that case our
generalization is equivalent to piecewise restricted Wheeler-Feynman
equations, otherwise completion by a finite action\cite{JMP2009} includes
other types of trajectories. In either case, along a piecewise defined
orbit, continuity of the spinor currents (\ref{mv1}) and (\ref{mt}) ensure
the well-posed continuation of the minimizer across each breaking point. For
example, solutions with continuous velocities yield trivially continuous
momenta (\ref{mv1}) and (\ref{mt}), so that any finite portion of a global
solution of the piecewise-restricted Wheeler-Feynman equations with
continuous velocities is a minimizer of the finite variational method with
suitable boundaries. For solutions with discontinuous velocities,
continuation to a global trajectory is possible using the (discontinuous)
velocity determined by solving conditions (\ref{mv1}) and (\ref{mt}) for the
most advanced velocity (to be shown elsewhere).

\section{\protect\bigskip Short-range orbits}

Wheeler-Feynman electrodynamics is a theory of direct interaction between
charges \cite{Fey-Whe, Darryl}. The boundary-value-variational-method
(previous section and Fig. 1) defines minimizers with a vanishing Frech\'{e}%
t derivative (\ref{Frechet}) between the time-spans of Fig. 1, rather than
globally-defined trajectories. As shown in Ref. \cite{JMP2009}, the
two-body-Euler-Lagrange equations can be cast in the form of Newtonian
equations of motion with each acceleration multiplied by the mass on the
left-hand-side, while the right-hand-side has the form of a Lorentz
-force-law. It is precisely these Euler-Lagrange-equations that define the
electromagnetic fields of Wheeler-Feynman theory as \emph{derived}
quantities evaluated \emph{on} trajectories.

Extending these fields defined by the
Lorentz-sector-of-Euler-Lagrange-equations to fields on positions outside
trajectories is tricky, because in a theory of trajectories one should: (i)
add a third particle to the variational problem; and (ii) arrange things
such that the third trajectory passes by the desired point. Obviously, a
third charge changes the minimization problem and perturbs the original
two-body-orbit, unless it can be placed so far that its couplings to the
original two-body-orbit are small. A bounded GAH two-body-orbit is special
because its far-fields vanish almost everywhere and a third trajectory can
be placed reasonably near without disturbing the two-body-orbit. Keeping in
mind that the far-fields are the strongest couplings to a third charge, we
investigate the existence of such (localized) short-range-two-body-orbits
(GAH).

We now adopt a unit system in which the speed of light is $c=1$ and apply
the usual formulae of electrodynamics to piecewise-defined-trajectories with
the exception of points where past/future velocities/accelerations are
undefined, i.e., the fields are undefined on a set of measure zero. We
consider continuous piecewise $C^{2}$ trajectories $\mathbf{x}_{k}(t_{k})$
enclosed by a sphere of radius $R$ in an inertial frame. We specify
space-time points $(t,R\mathbf{n})$ on the sphere by a time $t$ and unit
vector $\mathbf{n}$ normal to the surface of the sphere, and introduce an
index $k=1,2$ to label the charges.

The far-electric field of a point charge in the Wheeler-Feynman
electrodynamics is the sum of the half-advanced/half-retarded fields \cite%
{Fey-Whe}, 
\begin{equation}
\boldsymbol{E}(t,R\mathbf{n})=\frac{1}{2}\mathbf{E}^{adv}+\frac{1}{2}\mathbf{%
E}^{ret},  \label{genE}
\end{equation}%
while the far-magnetic field is given by 
\begin{equation}
\boldsymbol{B}(t,R\mathbf{n})=\frac{1}{2}\boldsymbol{n}_{+}\times \mathbf{E}%
^{adv}-\frac{1}{2}\boldsymbol{n}_{-}\times \mathbf{E}^{ret}.  \label{genB}
\end{equation}%
The unit vectors $\boldsymbol{n}_{\pm }$ point respectively from the
charge's advanced/retarded position to the position $R\mathbf{n}$ on the
sphere \cite{Jackson}. Trajectories are assumed to be bounded such that $|%
\mathbf{x}_{k}(t_{k})|<<R$, so that for each charge we have $\boldsymbol{n}%
_{+} \simeq
\boldsymbol{n}_{-}\equiv \boldsymbol{n}$.

The retarded far-electric and far-magnetic fields of a charge $q_{k}$ at the
space-time point $(t,R\mathbf{n})$ are piecewise-defined by the Li\'{e}%
nard-Wiechert formulas \cite{Jackson}

\begin{equation}
\mathbf{E}_{k}^{ret}(t,\mathbf{n})=\frac{q_{k}}{R}\frac{\mathbf{n}\times
\lbrack (\mathbf{n}-\mathbf{\mathbf{v}}_{k}\mathbf{)}\times \mathbf{a}_{k}]}{%
(1-\mathbf{n}\cdot \mathbf{v}_{k})^{3}},  \label{far-electric1}
\end{equation}%
and

\begin{equation}
\mathbf{B}_{k}^{ret}(t,\mathbf{n})=\mathbf{n}\times \mathbf{E}_{k}^{ret}(t,%
\mathbf{n}).  \label{far-magnetic1}
\end{equation}%
In Eq. (\ref{far-electric1}) we have used the far-field-limit in which the
light cone distance 
\begin{equation}
r_{k}(t_{k})\equiv |\mathbf{x}_{k}(t_{k})-R\mathbf{n|},  \label{lightdist}
\end{equation}%
is equal to $R$ since $|\mathbf{x}_{k}(t_{k})|<<R$. In Eq. (\ref%
{far-electric1}) $\mathbf{v}_{k}\equiv d\mathbf{x}_{k}/dt_{k}|_{t_{k}}$ and $%
\mathbf{a}_{k}$ $\equiv d^{2}\mathbf{x}_{k}/dt_{k}^{2}|_{t_{k}}$ are
respectively the charge's velocity and charge's acceleration at the retarded
time $t_{k}$ defined implicitly and piecewise by the \emph{retardation
condition}%
\begin{equation}
t_{k}=t-|\mathbf{x}_{k}(t_{k})-R\mathbf{n|},  \label{lightcone1}
\end{equation}%
where $|\cdot |$ denotes Cartesian distance. Equation (\ref{cone1}) defines $%
t_{k}$ as an implicit function of time $t$ with a piecewise defined
derivative 
\begin{equation}
\frac{dt_{k}}{dt}=\frac{1}{(1-\mathbf{n}\cdot \mathbf{v}_{k})}.
\label{dt1dt}
\end{equation}

Using Eq. (\ref{far-electric1}) to evaluate the far-magnetic field (\ref%
{far-magnetic1}) yields%
\begin{equation}
\mathbf{B}_{k}^{ret}(t,\mathbf{n})=-\frac{q_{k}\mathbf{n}}{R}\times \lbrack 
\frac{\mathbf{a}_{k}}{(1-\mathbf{n}\cdot \mathbf{v}_{k})^{2}}+\frac{(\mathbf{%
n}\cdot \mathbf{a}_{k})\mathbf{v}_{k}}{(1-\mathbf{n}\cdot \mathbf{v}_{k})^{3}%
}].  \label{newB}
\end{equation}%
The trajectory $\mathbf{x}_{k}(t_{k})$ is a function of $t$ from Eq. (\ref%
{lightcone1}) so using the chain rule and Eq. (\ref{dt1dt}) twice we can
re-write the far-magnetic-field (\ref{newB}) as 
\begin{equation}
\mathbf{B}_{k}^{ret}(t,\mathbf{n})=-\frac{q_{k}\mathbf{n}}{R}\times \frac{%
d^{2}}{dt^{2}}[\mathbf{x}_{k}(t_{k})].  \label{farman1}
\end{equation}%
The far-electric field is a linear function of the far-magnetic field
obtained using Eq. (\ref{far-magnetic1}) and the transversality property $%
\mathbf{n}\cdot \mathbf{E}_{k}^{ret}(t,\mathbf{n})=0$ of the far-electric
field (\ref{far-electric1}), i.e., 
\begin{equation}
\mathbf{E}_{k}^{ret}(t,\mathbf{n})=-\mathbf{n}\times \mathbf{B}_{k}^{ret}(t,%
\mathbf{n}).  \label{reconstruct}
\end{equation}%
In view of Eq. (\ref{reconstruct}), it suffices to study the vanishing of
the retarded-far-magnetic-fields. We further assume a symmetry that the
time-reversed orbit yields the same orbit rotated about an axis. For these
reverse--rotate-symmetric orbits the vanishing of the retarded far-fields
implies the vanishing of the advanced far-fields.

From now on charge $1$ is taken to be positive and equal to $q$ while charge 
$2$ is negative and equal to $-q$. The GAH along a bounded piecewise $C^{2}$
orbit is then expressed almost everywhere by 
\begin{equation}
\mathbf{B}^{ret}=\mathbf{B}_{1}^{ret}+\mathbf{B}_{2}^{ret}=-\frac{q\mathbf{n}%
}{R}\times \frac{d^{2}}{dt^{2}}(\mathbf{x}_{1}(t_{1})-\mathbf{x}%
_{2}(t_{2}))=0.  \label{vanishing}
\end{equation}%
In the family of orbits with discontinuous velocities one can readily
construct bounded orbits with vanishing far-fields; e.g.
piecewise-constant-velocity orbits with trajectories consisting of polygonal
lines. These are bounded orbits with each acceleration vanishing piecewise,
so that the radiation fields vanish. The question that needs to be answered
is ``Do we need these velocity discontinuities?".

Equation (\ref{vanishing}) is a (linear) \emph{neutral differential delay
equation} with piecewise-linear continuous solutions defined on the
intervals $t \in (t_{\sigma -1}^{+},t_{\sigma }^{-})$, with $\sigma \in 
\mathbb{Z}
$ by 
\begin{equation}
\mathbf{x}_{1}(t_{1})-\mathbf{x}_{2}(t_{2})=\mathbf{D}_{\sigma }(\mathbf{n})+%
\mathbf{n}f_{\sigma }(t,\mathbf{n})+(t-t_{\sigma }^{-})\boldsymbol{V}%
_{\sigma }(\mathbf{n}),  \label{dipole}
\end{equation}%
where the $\mathbf{D}_{\sigma }(\mathbf{n})$ and $\boldsymbol{V}_{\sigma }(%
\mathbf{n})$ \ are arbitrary bounded functions and the $\ f_{\sigma }(t,%
\mathbf{n})$ are bounded and piecewise $C^{2}$. It is possible to choose $%
\mathbf{n}\cdot \mathbf{D}_{\sigma }(\mathbf{n})\mathbf{=0}$ and adjust $%
\mathbf{D}_{\sigma }(\mathbf{n})$ in each interval to make the
left-hand-side of Eq. (\ref{dipole}) continuous.

Along a spatially bounded orbit, Eq. (\ref{lightcone1}) is approximated for
large values of $R$ by 
\begin{equation}
t_{k}\mathbf{=}t-R+\mathbf{n}\cdot \mathbf{x}_{k}(t_{k}).  \label{cone1}
\end{equation}%
Notice that Eqs. (\ref{cone1}) yield an implicit relation between $t_{1}$
and $t_{2}$, 
\begin{equation}
t_{1}-t_{2}=\mathbf{n}\cdot (\mathbf{x}_{1}(t_{1})-\mathbf{x}_{2}(t_{2})).
\label{t1t2}
\end{equation}%
Given the trajectories $\mathbf{x}_{1}(t_{1})$ and $\mathbf{x}_{2}(t_{2})$,
Eq. (\ref{t1t2}) and the implicit function theorem yield $t_{1}$ as a
function of $t_{2}$ and $\mathbf{n}$. Define the \emph{influence interval }%
of point $(t_{2},\mathbf{x}_{2}(t_{2}))$ by the interval containing $t_{1}$
when $\mathbf{n}$ varies arbitrarily in Equation (\ref{t1t2}), i.e., 
\begin{equation}
t_{2}-|\mathbf{x}_{1}(t_{1})-\mathbf{x}_{2}(t_{2})|<t_{1}<t_{2}+|\mathbf{x}%
_{1}(t_{1})-\mathbf{x}_{2}(t_{2})|.  \label{tspan}
\end{equation}

The time span (\ref{tspan}) is from the retarded lightcone-time of $(t_{2},%
\mathbf{x}_{2}(t_{2}))$ to the advanced lightcone-time of $(t_{2},\mathbf{x}%
_{2}(t_{2}))$, as along the sewing chain illustrated in Fig. 1. Notice that
the future lightcone appeared naturally in the two-particle problem, even
though we were dealing only with the retardation conditions (\ref{cone1}).
It follows from Eqs. (\ref{t1t2}) and (\ref{dipole}) that%
\begin{equation}
f_{\sigma }(t,\mathbf{n})=(t_{1}-t_{2})-(t-t_{\sigma }^{-})\mathbf{n}\cdot
V_{\sigma }(\mathbf{n}).  \label{newf}
\end{equation}%
and we can therefore re-write Eq. (\ref{dipole}) as%
\begin{equation}
\mathbf{x}_{1}(t_{1})-\mathbf{x}_{2}(t_{2})=\mathbf{D}_{\sigma }(\mathbf{n}%
)+(t_{1}-t_{2})\mathbf{n}-(t-t_{\sigma }^{-})\mathbf{n}\times \mathbf{L}%
_{\sigma }(\mathbf{n}).  \label{newdipole}
\end{equation}%
where $\mathbf{L}_{\sigma }(\mathbf{n})\equiv \mathbf{n\times }V_{\sigma }(%
\mathbf{n})$. Since linear growth in a constant direction is unbounded, the
only globally $C^{2}$ orbit must have $\boldsymbol{L}_{\sigma }(\mathbf{n}%
)=0 $ $\forall \sigma $, and it follows from Eq. (\ref{newdipole}) with $%
\boldsymbol{L}_{\sigma }(\mathbf{n})=0$ that 
\begin{equation}
\mathbf{x}_{1}(t_{1})-\mathbf{x}_{2}(t_{2})=\mathbf{D}_{\sigma }(\mathbf{n}%
)+(t_{1}-t_{2})\mathbf{n}.  \label{express}
\end{equation}%
The derivative of Eq. (\ref{express}) respect to time yields 
\begin{equation}
\frac{\mathbf{v}_{1}}{(1-\mathbf{n}\cdot \mathbf{v}_{1})}-\frac{\mathbf{v}%
_{2}}{(1-\mathbf{n}\cdot \mathbf{v}_{2})}=K_{12}\mathbf{n},  \label{velocity}
\end{equation}%
where%
\begin{equation}
K_{12}=\frac{1}{(1-\mathbf{n}\cdot \mathbf{v}_{1})}-\frac{1}{(1-\mathbf{n}%
\cdot \mathbf{v}_{2})}.  \label{defK}
\end{equation}

Equation (\ref{t1t2}) allow us to move $\mathbf{n}$ in a cone with axis
along $\mathbf{x}_{1}(t_{1})-\mathbf{x}_{2}(t_{2})\neq 0$ in a way that
fixes $t_{1}$ and $t_{2}$ while changing $t$ with Eqs. (\ref{cone1}).\ On
the other hand, for fixed $t_{1}$ and $t_{2}$ the left-hand-side of Eq. (\ref%
{velocity}) spans a plane of the \emph{fixed} vectors $\mathbf{v}_{1}(t_{1})$
and $\mathbf{v}_{2}(t_{2})$, so that Eq. (\ref{velocity}) can hold only if $%
K_{12}=0$, which combined with Eqs. (\ref{velocity}) and (\ref{defK}) yields 
\begin{equation}
\mathbf{v}_{1}(t_{1})=\mathbf{v}_{2}(t_{2}).  \label{rest12}
\end{equation}%
Equation (\ref{rest12}) defines globally-constant velocities along a fixed
direction, which in turn implies unbounded motion unless $\mathbf{v}_{1}=%
\mathbf{v}_{2}=0$, as discussed in Ref. \cite{Marino}. This impossibility
follows if velocities are to be continuous.

Nontrivial alternatives to this unsatisfactory conclusion necessitate the
introduction of discontinuities by varying the direction of the
piecewise-velocity-like term $\boldsymbol{L}_{\sigma }(\mathbf{n})\neq 0$ of
Eq. (\ref{newdipole}) in each interval. The piecewise derivative of Eq. (\ref%
{newdipole}) respect to time yields%
\begin{equation}
\frac{\mathbf{v}_{1}}{(1-\mathbf{n}\cdot \mathbf{v}_{1})}-\frac{\mathbf{v}%
_{2}}{(1-\mathbf{n}\cdot \mathbf{v}_{2})}=K_{12}\mathbf{n}-\mathbf{n}\times 
\mathbf{L}_{\sigma }(\mathbf{n}).  \label{novoplane}
\end{equation}%
Notice that $K_{12}$ is still given by Eq. (\ref{defK}) and with nonzero $%
\boldsymbol{L}_{\sigma }(\mathbf{n})$ the right-hand-side of Eq. (\ref%
{novoplane}) forms a complete 3-dimensional basis to express any vector
(inside or outside the plane of $\mathbf{v}_{1}(t_{1})$ and $\mathbf{v}%
_{2}(t_{2})$ ). Equation (\ref{t1t2}) still allows one to move $\mathbf{n}$
in a cone with axis along $\mathbf{x}_{1}(t_{1})-\mathbf{x}_{2}(t_{2})\neq 0$
in a way that fixes $t_{1}$ and $t_{2}$ while $t$ \ changes with Eqs. (\ref%
{cone1}). By choosing $K_{12}$ and a nonzero $\boldsymbol{L}_{\sigma }(%
\mathbf{n})$ for each $t\in $ $(t_{\sigma -1}^{+},t_{\sigma }^{-})$ we can
describe any vector on the left-hand-side of Eq. (\ref{novoplane}), so that
there is no inconsistency.

As an example, time-reversible orbits satisfying Eq. (\ref{novoplane}) are
piecewise-constant-velocity orbits generated by having one velocity jump at
a given time while the other velocity jumps either in the backward or
forward lightcone-times symmetrically, as well as at every time in the
forward and backward light-cones of a discontinuity time (the sewing chain
illustrated in Fig. 1). These piecewise-linear polygonal orbits can be shown
to satisfy Eq. (\ref{vanishing}) by direct substitution and use of Eq. (\ref%
{dt1dt}). \ In the following we show that Eq. (\ref{newdipole}) and the
implicit function theorem yield a consistent piecewise-defined trajectory $%
\mathbf{x}_{1}(t_{1})$ from a given piecewise-defined trajectory $\mathbf{x}%
_{2}(t_{2})$.

Notice that for given continuous and piecewise $C^{1}$ $\mathbf{x}%
_{2}(t_{2}) $, $\mathbf{D}_{\sigma }(\mathbf{n})$ and $\boldsymbol{L}%
_{\sigma }(\mathbf{n})$, in general Eq. (\ref{newdipole}) determines only a
function $\mathbf{x}_{1}(t_{1},\mathbf{n})$ of the \emph{two variables} $%
(t_{1},\mathbf{n})$ through 
\begin{equation}
\mathbf{x}_{1}(t_{1},\mathbf{n})=\mathbf{x}_{2}(t_{2})+\mathbf{D}_{\sigma }(%
\mathbf{n})+(t_{1}-t_{2})\mathbf{n}-(t-t_{\sigma }^{-})\mathbf{n}\times 
\mathbf{L}_{\sigma }(\mathbf{n}).  \label{x1t1n}
\end{equation}%
The implicit function theorem further determines $t_{2}$ and $t$ as
functions of $t_{1}$ and $\mathbf{n}$ from Eqs. (\ref{cone1}), (\ref{t1t2})
and (\ref{x1t1n}). For the implicit function theorem to yield a consistent
trajectory, we must satisfy the consistency requirement that $\mathbf{x}%
_{1}(t_{1},\mathbf{n})$ determined by Eq. (\ref{x1t1n}) is a function of $%
t_{1}$ only, i.e., 
\begin{equation}
\frac{\partial \mathbf{x}_{1}(t_{1},\mathbf{n})}{\partial \mathbf{n}}=0.
\label{consistency}
\end{equation}

Condition (\ref{consistency}) applied to the right-hand-side of Eq. (\ref%
{x1t1n}) is the extra condition determining a consistent trajectory. Since
condition (\ref{consistency}) must hold for all values of $t_{1}$ in each
piecewise interval of the orbit, we must also have inside each piecewise
interval that 
\begin{equation}
\frac{\partial ^{2}\mathbf{x}_{1}(t_{1},\mathbf{n})}{\partial t_{1}\partial 
\mathbf{n}}=\frac{\partial }{\partial \mathbf{n}}(\frac{\partial \mathbf{x}%
_{1}(t_{1},\mathbf{n})}{\partial t_{1}})=0,  \label{mixed}
\end{equation}%
which can be expressed as%
\begin{equation}
\frac{\partial }{\partial \mathbf{n}}[(\mathbf{v}_{2}-\mathbf{n})\frac{%
\partial t_{2}(t_{1},\mathbf{n})}{\partial t_{1}}+\mathbf{n}-\frac{\partial
t(t_{1},\mathbf{n})}{\partial t_{1}}\mathbf{n}\times \boldsymbol{L}_{\sigma
}(\mathbf{n})]=0.  \label{compatibility}
\end{equation}%
The general solution to Eq. (\ref{compatibility}) involves an arbitrary
piecewise-defined function $A_{\sigma }(t_{1})$, i.e., 
\begin{equation}
\mathbf{n+}\frac{\partial t_{2}(t_{1},\mathbf{n})}{\partial t_{1}}(\mathbf{v}%
_{2}-\mathbf{n})-\frac{\partial t(t_{1},\mathbf{n})}{\partial t_{1}}\mathbf{n%
}\times \boldsymbol{L}_{\sigma }(\mathbf{n})=A_{\sigma }(t_{1}).
\label{generalt1}
\end{equation}%
A symmetric condition follows by exchanging indices $1$ and $2$ in Eq.(\ref%
{compatibility}), introducing an arbitrary $B_{\sigma }(t_{2})$ and changing
the sign of $\boldsymbol{L}_{\sigma }(\mathbf{n})$, yielding 
\begin{equation}
\mathbf{n+}\frac{\partial t_{1}(t_{2},\mathbf{n})}{\partial t_{2}}(\mathbf{v}%
_{1}-\mathbf{n})+\frac{\partial t(t_{2},\mathbf{n})}{\partial t_{2}}\mathbf{n%
}\times \boldsymbol{L}_{\sigma }(\mathbf{n})=B_{\sigma }(t_{2}).
\label{generalt2}
\end{equation}%
The partial derivatives in Eqs. (\ref{generalt1}) and (\ref{generalt2}) can
be evaluated using Eqs. (\ref{cone1}), yielding 
\begin{eqnarray}
\frac{A_{\sigma }(t_{1})}{(1-\mathbf{n}\cdot \mathbf{v}_{1})}-\frac{\mathbf{v%
}_{2}(t_{2})}{(1-\mathbf{n}\cdot \mathbf{v}_{2})} &=&K_{12}\mathbf{n}-%
\mathbf{n}\times \mathbf{L}_{\sigma }(\mathbf{n}),  \label{final1} \\
\frac{\mathbf{v}_{1}(t_{1})}{(1-\mathbf{n}\cdot \mathbf{v}_{1})}-\frac{%
B_{\sigma }(t_{2})}{(1-\mathbf{n}\cdot \mathbf{v}_{2})} &=&K_{12}\mathbf{n}-%
\mathbf{n}\times \mathbf{L}_{\sigma }(\mathbf{n}),  \label{final2}
\end{eqnarray}%
where again $K_{12}$ is defined by Eq. (\ref{defK}).

Notice that Eqs. (\ref{final1}) and (\ref{final2}) with $A_{\sigma }(t_{1})=%
\mathbf{v}_{1}(t_{1})$ and $B_{\sigma }(t_{2})=\mathbf{v}_{2}(t_{2})$ bring
back Eq. (\ref{novoplane}) as the single necessary condition to construct a
consistent piecewise $C^{1}$ continuous trajectory $\mathbf{x}_{1}(t_{1})$
from a given piecewise $C^{1}$ continuous trajectory $\mathbf{x}_{2}(t_{2})$
by the implicit function theorem. We stress that velocity discontinuities
are \textit{absolutely necessary}. The discontinuities introduced by the
nonzero $\boldsymbol{L}_{\sigma }(\mathbf{n})$ in each piecewise interval
play an essential role in the solution as discussed below Eq. (\ref%
{novoplane}). Trying to solve either Eq. (\ref{final1}) or Eq. (\ref{final2}%
) with $\boldsymbol{L}_{\sigma }(\mathbf{n})=0$ stumbles with the former
obstruction that rotation of $\mathbf{n}$ with fixed $t_{1}$ and $t_{2}$
places $\mathbf{n}$ either outside the plane of $A_{\sigma }(t_{1})$ and $%
\mathbf{v}_{2}(t_{2})$ or outside the plane of $B_{\sigma }(t_{2})$ and $%
\mathbf{v}_{1}(t_{1})$. The nonzero $\boldsymbol{L}_{\sigma }(\mathbf{n})$
provides the third linearly independent direction forming a complete basis
to express $\mathbf{n}$ in Eqs. (\ref{final1}) and (\ref{final2}).

It can be seen that piecewise-constant-velocity polygonal orbits can be
constructed using the suitable $\mathbf{L}_{\sigma }(\mathbf{n})$ defined by
Eq. (\ref{novoplane}), after which the implicit function theorem constructs
consistent continuous piecewise $C^{1}$ trajectories. It would be desirable
to find bounded minimizer orbits of the variational method \cite{JMP2009}
satisfying the vanishing far-field conditions (\ref{final1}) and (\ref%
{final2}), as first conjectured for the orbits studied in Ref. \cite%
{stiff-hydrogen}. The justification to generalize to trajectories with
discontinuous derivatives is to include short-range bounded GAH orbits in
the family of physically possible orbits. This was used in Ref. \cite%
{stiff-hydrogen} to predict spectroscopic lines of hydrogen within a few
percent agreement with the predictions of quantum mechanics.

\section{Discussion and Conclusion}

The fact that accelerations are discontinuous is expected because the
Wheeler-Feynman equations of motion are explicitly neutral for the
accelerations. Consequently, it could seem that a theory of
piecewise-restricted Wheeler-Feynman equations of motion should have only
acceleration discontinuities, a fact that already introduces discontinuous
fields \emph{and} demands a generalization of electrodynamics. We have seen
that generalizing to trajectories with discontinuous accelerations is not
sufficient to include bounded two-body orbits with vanishing far-fields. Our
analysis starting from the variational method as the fundamental principle
has shown that, in general, the velocities are also expected to be
discontinuous at the same "generalized breaking points" along the minimizer
orbits. Our analysis, using the variational method as a boundary-value
problem, shows that the most general solution of the Wheeler-Feynman neutral
differential delay equations has discontinuous accelerations \emph{and}
velocities.

The form of Eq. (\ref{mt}) is reminiscent of the energy operator used
formally in quantum mechanics. Since the evolution parameter is arbitrary,
the same parameter can be used in a Lorentz-transformed frame, such that the
momentum currents transform by like a four-vector. The existence of four
components that must be continuous and given by the partial derivatives of a
scalar invariant is again analogous to the quantum Dirac equation and
suggests a property analogous to spin for the point charges. It is
remarkable that Wheeler-Feynman electrodynamics completed with a finite
action endows the point charges with a spin-like property. The existence of
a spinorial four-component momentum current (Eqs. (\ref{mv1}) and (\ref{mt}%
)) is due to the parametrization-invariance symmetry of the electromagnetic
variational method \cite{JMP2009}. Otherwise a generic action with delayed
interaction has only three momentum currents continuous at breaking points
of minimizer orbits.

In Ref. \cite{Driver} only globally $C^{2}$ solutions were sought for the
seemingly non-neutral one dimensional motion, so that piecewise-defined
solutions with discontinuous velocities awaited study. In considering this,
it is important that the electromagnetic-action-functional (\ref{action1}) of
Ref. \cite{JMP2009} yields a \emph{neutral-boundary-value-variational-method}%
, as opposed to the non-neutral variational methods of classical mechanics.

The variational principles of classical mechanics are two-point boundary
value problems that are equivalent to an initial-value problem with initial
velocity chosen to hit the final trajectory point. Moreover, in the
classical problem there is no issue of velocity continuity, because the
``history" for a finite-dimensional ODE is a point. On the contrary, for the
electromagnetic variational method, choosing the ``initial velocity" to
shoot the final point either requires a velocity discontinuous with the past
boundary history \emph{or} the trajectory arrives at the final point with
the wrong velocity. The generalization to discontinuous velocities extends
the solvability of the electromagnetic-boundary-value-problem to a larger
class of boundary value data, which is the second advantage of our extension
of Wheeler-Feynman electrodynamics.

Electromagnetism was originally formulated with the integral laws of Ampere,
Gauss and Faraday, and only much later differential equations holding
everywhere were introduced by Maxwell. The requirement of a second
derivative existing everywhere is actually not needed\ for particle
dynamics, where one is concerned only with the integral of the force along
the trajectory. The variational method is a step back from Maxwell's
equations in the sense of weak solutions. Replacing Maxwell's equations by
the vanishing of the Frech\'{e}t derivative (\ref{Frechet}) along a
continuous trajectory with boundaries in future and past yields solutions
defined only on bounded time-intervals. From these segments of orbits one
can construct the fields as \emph{derived quantities }\cite{Fey-Whe,Darryl}.
Since fields constructed in this fashion involve a retarded and a advanced
position, before defining fields everywhere in space we need to extend to a
global trajectory. Extension is possible using conditions (\ref{mv1}) and (%
\ref{mt}) and in general involves velocity and acceleration discontinuities.
It is then tempting to translate our generalized electrodynamic quantities
into the concepts of Maxwell's electrodynamics, but it must be done
carefully; The derivation of the differential form of Maxwell's equations
given in \cite{Fey-Whe} holds only in regions where the extended fields are $%
C^{1}$. Consequently, Poynting's theorem in integral form holds in regions
where the fields are $C^{1}$. Nevertheless, quantities like energy flux and
field energy in a volume have a meaning even for discontinuous fields, so
that a statistical interpretation could be sought in the case of
discontinuous fields; For example, the generalized flux of the Poynting
vector is an integral that can be evaluated even when fields are undefined
on sets of measure zero. The Poynting vector $\boldsymbol{P}=\boldsymbol{E}%
\times \boldsymbol{B}$ evaluated with Eqs. (\ref{genE}) and (\ref{genB}) at $%
(t,R\mathbf{n})$ becomes 
\begin{equation}
\boldsymbol{P=}\frac{1}{4}\{|\mathbf{E}^{adv}|^{2}-|\mathbf{E}^{ret}|^{2}\}%
\boldsymbol{n}  \label{PoyntingX}
\end{equation}%
where single bars denote the Euclidean modulus.

We stress that a condition of non-radiation is weaker than our GAH, as
follows; The GAH \emph{implies} the vanishing of the flux (\ref{PoyntingX})
because $|\mathbf{E}^{ret}|^{2}=|\mathbf{E}^{adv}|^{2}=0$ \emph{almost
everywhere}, while the converse is not true, i.e., the vanishing of the flux
integral alone does not imply the GAH. \ For example, the circular orbits of
Refs. \cite{Schoenberg,Schild} do not satisfy the GAH, and even though these
orbits do not radiate on average, circular orbits have non-vanishing
far-fields to disturb a ``third charge of the variational method" (i.e., are
not short-range). In Ref. \cite{Schild} a model for the neutron was
attempted, and even if it had not failed for other reasons, it would yield a
neutron with far-fields. As regards non-vanishing far-fields, a first
attempt to overcame the GAH-deficiency of circular orbits \cite{Schild} was
the perturbation theory of Ref. \cite{stiff-hydrogen} that added
high-frequency modes of the tangent dynamics to enforce Eq. (\ref{vanishing}%
) at the frequency of the circular orbit.

We have shown that our variational method yields a dynamical system even for
limiting orbits with discontinuous velocities. For example, along
piecewise-constant-velocity polygonal orbits the variational equations of
motion would be applied as follows; On discontinuity corner points with no
acceleration defined, one enforces continuity of momentum only, Eq. (\ref%
{mv1}). At other points, wherever accelerations \emph{are} defined, one uses
the usual Wheeler-Feynman equations of motion. (Notice that
piecewise-constant-velocity polygonal orbits have vanishing far-fields but
obviously do not satisfy the equations of motion, unless charges are far
apart).

We have demonstrated five different and important reasons to study orbits
with discontinuous derivatives: (i) inclusion of bounded GAH orbits as
short-range orbits; (ii) compatibility with the conservation of Noether%
\'{}%
s momentum; (iii) compatibility with the neutrality of the equations of
motion of the Wheeler-Feynman electrodynamics; (iv) the fact that the
variational method is natural in a space completed to contain orbits with
discontinuous velocities; and (v) inclusion of limits of sequences of
non-radiating orbits.

The physical need for trajectories with discontinuous velocities is
justified as limiting orbits defined by Cauchy sequences of bounded orbits
which \emph{must} develop kinks in the short-range limit (i.e., the GAH). In
Ref. \cite{stiff-hydrogen} the GAH-deficiency of circular orbits was removed
with a perturbative Fourier series that solved Eq. (\ref{vanishing}) at the
first harmonic frequency only. As we have shown here, the perturbative
series of Ref. \cite{stiff-hydrogen} should converge to an orbit with \emph{%
discontinuous velocities}. The short-range condition of Ref. \cite%
{stiff-hydrogen} predicted orbits and spectral lines in the atomic magnitude
with a surprising precision, so that we can claim agreement with experiment
and quantum mechanics. Piecewise-defined minimizers have also been used
successfully to explain double-slit diffraction in Ref. \cite{double-slit}.

From our generalized electrodynamics with discontinuous derivatives it
should be possible to derive a generalized electrodynamics with delayed-only
interactions and self-interaction, using the generalized absorber hypothesis
(GAH) in close analogy with the derivation of Wheeler and Feynman \cite%
{Fey-Whe}. Since we expect solutions with velocity discontinuities, Taylor
expansions of deviating arguments should be avoided or piecewise-restricted.
It is known that many-component delay differential equations behave like
neutral differential delay equations when some solution components are
discontinuous at breaking points, in the sense that the discontinuous
derivatives never smooth out \cite{ChrisPaul}. Therefore, the third
derivative should be generalized by restricting it to a left-derivative and
a right-derivative at breaking points. Also, the generalized absorber
hypothesis with discontinuous fields no longer implies the vanishing of the
difference of retarded and advanced universal fields everywhere, as used by
Wheeler and Feynman \cite{Fey-Whe, Narlikar}. Our Eq. (\ref{novoplane}) and
its advanced version give the corresponding weaker generalization to this
former stronger condition of a vanishing difference of retarded and advanced
fields everywhere. We speculate that Eq. (\ref{novoplane}) should be the
starting point for a generalized theory of self-interaction free of the
pervasive runaways of the two-body problem with the usual self-interaction.

Last, we speculate that \emph{unbounded} scattering orbits are different
from the bounded orbits studied here. Along unbounded orbits Eq. (\ref%
{dipole}) contains the extra secular term with a constant $\boldsymbol{V}%
_{\sigma }(\mathbf{n})\neq 0$. The dependence on boundary segments and time
separation must be investigated for scattering trajectories with
discontinuous velocities and accelerations at the boundaries; For example,
even if the history segment $(O^{-},$ $O^{+})$ of Figure 1 is assumed $%
C^{\infty }$, the forward sewing chain of $O^{+}$ places a breaking point $%
f_{3}$ in the history segment $(L^{-},$ $L^{+})$ of particle $1$. Unless
histories are very special so that derivatives are continuous at $O^{+}$, in
general the history $(L^{-},$ $L^{+})$ should involve a discontinuous
derivative at point $f_{3}$. Scattering trajectories are likely to have
future continuations involving stiffer jumps at later times, so that
particles collide with laboratory boundaries, which can be regarded as a
generalized type of radiative loss.

\bigskip

\subsection{Figure Captions}

\bigskip

Figure 1: Illustrated in red is the initial point $O_{A}$ of trajectory $1$
plus the segment of trajectory $2$ inside the lightcone of $O_{A}$, i.e.,
from point $O^{-}$ to point $O^{+}$ and \ the endpoint $L_{B}$ of trajectory 
$2$ plus the segment of trajectory of particle $1$ inside the lightcone of $%
L_{B}$, i.e., from point $L^{-}$ to point $L^{+}$. The trajectory of
particle $1$ of the variational method goes from $O_{A}$ to $L^{-}$ (blue
line) while the trajectory of particle $2$ goes from $O^{+}$ to $L_{B}$
(green line). The first breaking point is point $O^{+}$which generates a
forward sewing chain of breaking points $f_{1},f_{2,}f_{3}$ while endpoint $%
L^{-}$ is a breaking point generating a backward sewing chain of breaking
points $b_{1},b_{2},b_{3}$.

\end{document}